\documentclass{vgtc}        

\graphicspath{{figures/}{pictures/}{images/}{./}} 

\usepackage{times}

\vgtccategory{Research}
\vgtcpapertype{Application/Design Study}
\vgtcinsertpkg

\title{Integrating Annotations for Sonifications and Physicalizations}

\author{
Rhys Sorenson-Graff\thanks{e-mail: sorensor@whitman.edu}\\ %
        \scriptsize Whitman College %
\and
S. Sandra Bae\thanks{e-mail: sandra.bae@colorado.edu}\\ %
     \scriptsize CU Boulder %
\and
Jordan Wirfs-Brock\thanks{e-mail: wirfsbrj@whitman.edu}\\ %
     \scriptsize Whitman College
}

\abstract{
Annotations are a critical component of visualizations, helping viewers interpret the visual representation and highlighting critical data insights.
Despite their significant role, we lack an understanding of how annotations can be incorporated into other data representations, such as physicalizations and sonifications. 
Given the emergent nature of these representations, sonifications, and physicalizations lack formalized conventions (e.g., design space, vocabulary) that can introduce challenges for audiences to interpret the intended data encoding. 
To address this challenge, this work focuses on how annotations can be more tightly integrated into the design process of creating sonifications and physicalizations. In an exploratory study with 13 designers, we explore how visualization annotation techniques can be adapted to sonic and physical modalities. Our work highlights how annotations for sonification and physicalizations are inseparable from their data encodings. 
}


\teaser{
  \centering
  \includegraphics[width=0.9\linewidth, alt={A view of a city with buildings peeking out of the clouds.}]{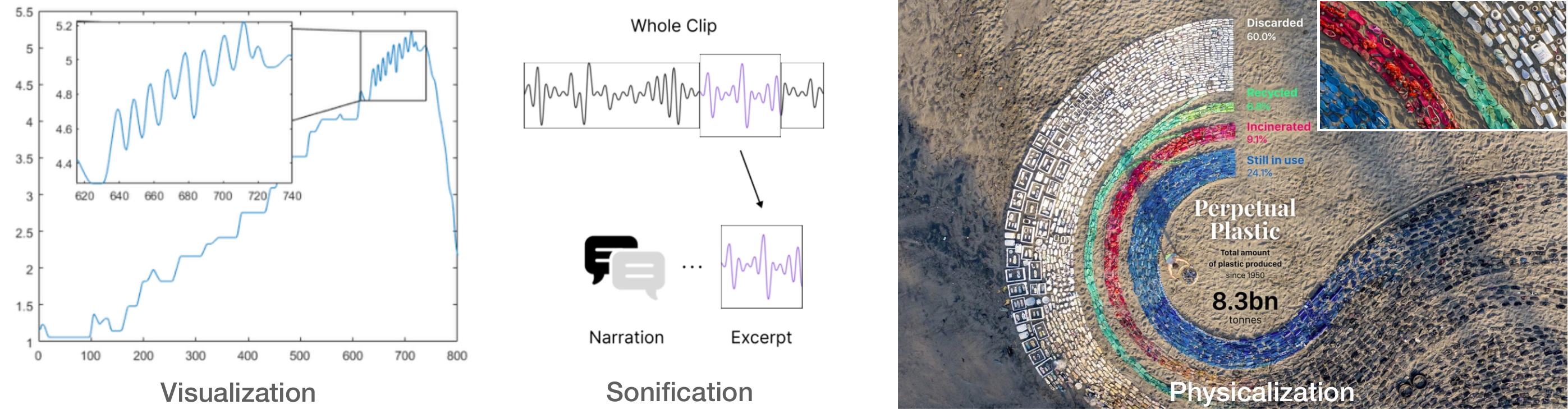}
  \caption{Examples of geometric annotations used in a visualization, sonification, and physicalization. Geometric annotations draw attention to a specific section of the data representation, providing additional context, detail, and clarity to a section if it contains crucial information or is of significant interest to the viewer~\cite{rahman2023qualitative}. Visualizations can integrate geometric annotations with call-out boxes. Sonifications can highlight specific excerpts using sub-clips of audio. Physicalizations can present multiple frames of reference to emphasize different
  perspectives that zoom in and out of the physicalization (photo credit to Klauss et al.~\cite{perpetual-plastic}).
  }
  \label{fig:teaser}
}

\renewcommand{\manuscriptnotetxt}{}
\nocopyrightspace

\graphicspath{{figs/}{figures/}{pictures/}{images/}{./}} 

\usepackage{tabu}                    
\usepackage{booktabs}                
\usepackage{lipsum}                   
\usepackage{mwe}                     
\usepackage{mathptmx}            
\usepackage{paralist}
\usepackage{array}
\usepackage{subfig}
\usepackage[font={scriptsize,sf},tableposition=top]{caption}
\usepackage{float}
\floatstyle{plaintop}
\restylefloat{table}
\usepackage{enumitem}
\usepackage{xcolor}
\usepackage{colortbl}
\usepackage{tabularx}
\usepackage[para,online,flushleft]{threeparttable}
\usepackage{cite,etoolbox}

\def\techname{\textit{SonNotate/PhysNotate}}

\begin{document}


\firstsection{Introduction}
\maketitle
Annotations---cues designed to direct a viewer's attention---are a ``vital component'' of data representations~\cite{rahman2023qualitative, heer2012interactive, zhao2017annotations}. 
They teach users how to interpret the visual representation, highlighting critical insights~\cite{rahman2023qualitative}, facilitating exploration~\cite{kang2014characterizing, sevastjanova2022visinreport, shrinivasan2009connecting, mahyar2012note} and even collaborative data analysis~\cite{chen2010click, chen2010touch}. 
Despite the significant role annotations serve in visualizations, we lack insight on how to incorporate annotations in \textit{other} data representations, such as physicalizations and sonifications. Rather than encoding data into visual marks and channels, physicalizations tangibly encode data into physical or geometric properties of materials~\cite{jansen2015opportunities} and sonifications encode data into sounds~\cite{kramer2000auditory}.
This wide cast of data encoding channels results in diverse artifacts and design possibilities~\cite{bae2022making, bae2022towards, bae2022exploring, jansen2015opportunities, hermann2011sonification}. However, a challenge with these emergent data representations and their data encoding processes is the lack of formalized conventions (e.g., design space, vocabulary) \cite{hermann2011sonification, hornecker2023design}.

\begin{table*}[t!]
\small
\begin{tabular}{
                >{\raggedright}
                p{1.5cm}
                >{\raggedright}
                p{5.5cm}
                >{\raggedright}
                p{3cm}
                >{\raggedright}
                p{3cm}
                >{\raggedright}
                p{3cm}}
 \toprule 
 \textbf{Technique} & 
 \textbf{Purpose} &
 \textbf{Visualization Example(s)} & \textbf{Sonification Example(s)} & \textbf{Physicalization Example(s)}\tabularnewline\arrayrulecolor{black!30}\midrule
    Text & Describing chart elements (e.g., indicate direct values from the dataset, provide additional context) 
    & 
    Written descriptions, value labels, legends & Speech in series or parallel with audio &
    Descriptions, value labels, legends 
    \tabularnewline\addlinespace[0.2em]

    Enclosure & Grouping related information to help interpret data by creating a full or partial boundary & Bounding shapes such as brackets, rectangles, ellipses 
    & A sonic element (e.g., undertone, reverb) lasting for a discrete amount of time
    & In 3D, objects such as boxes, jars, containers
\tabularnewline\addlinespace[0.2em]

    Connector & Establishing a visual connection between two areas of interest (e.g. between a data point and a text description, or between two data points) & 
    Pointer (directed or undirected) to a specific spot or value (e.g., line, arrow)  
    & 
    Verbal cue that calls attention to a specific value using narrative (e.g., ''this sound'' $<$\texttt{sound}$>$ ``means...'' ) &
    Pointer (directed or undirected) to a specific value connecting to data point(s) (e.g., 3D: Tube, wire, cord)\tabularnewline\addlinespace[0.2em]
   
    Mark & Acting as visual indicators in different chart types&
    Symbols (e.g., stars, circles)  placed next to a particular data point  
    &
    An auditory icon that marks a relevant data point (e.g., a bell at the maximum value) &
    Objects placed next to a particular data point  (e.g., lights, pillars)   \tabularnewline\addlinespace[0.2em]

   Color & Visually distinguishing between different categories or data points and drawing attention to specific elements or regions of a visualization & 
   Variations in color (e.g., hue, saturation) & 
   Variations in the quality of the sounds (e.g., timbre, volume) &
   Variations in texture (e.g., roughness, smoothness) or temperature \tabularnewline\addlinespace[0.2em]

  Trend & Depicting changes along a particular axis. Can be in direction (e.g., upward or downward),  in magnitude (e.g., rapid or slower), or correlation relationship (e.g., positive or negative correlation). & 
  Line that summarizes statistical properties of the data or subset of the data & 
  A separate summary sound (in addition to discrete sounds representing individual data points) &
  Using a person's body and perspective to explore trends (walk from the min to the max of the data values). \tabularnewline\addlinespace[0.2em]
    
   Geometric & Highlight and transform a particular subset of the data& 
   A call-out box that zooms in a portion of the data & 
   Excerpting and transforming a subset of the sound (e.g., playing a clip slower) & 
   A viewer walking around an object to change focus using size and perspective \tabularnewline
   
\bottomrule
\end{tabular}
\caption{Summary of visual annotation techniques outlined in Rahman et al.~\cite{rahman2023qualitative}, extrapolated to physicalizations and sonifications. Examples of annotation techniques for sonifications and physicalizations are identified from close readings of case studies (see \autoref{sec3:design-process}).}
\label{tab:annotation}
\vspace{-0.4cm}
\end{table*}

Visualization designers can rely on the audiences' collective knowledge of how to interpret conventional graphs (e.g., scatterplots, bar charts), but this assumption is not necessarily true for sonifications and physicalizations. Consequently, we assert that annotations are critical to communicating the data mapping behind these emergent data representations. Yet, from our observations and personal experiences of designing physicalizations and sonifications, we noticed that the typical process of designing these emergent data representations often treats annotations as a separate and last layer to add to the data representations. Only later, through testing and refinement, do data designers consider how to communicate their data representation to stakeholders with annotations. We explore how annotations can be an integral part of a data representation by considering the following research provocation: \textit{How might we design more understandable
data representations by inverting the typical design process so that it centers annotations?}

To explore this provocation, we created \techname, a card-based game, as a design probe \cite{gaver1999design, boehner2007hci} on how we might support physicalization and sonification designers in taking a more annotation-centric approach. Using a research-through-design (RTD) approach~\cite{2007zimmerman, sedlmair2012design}, this probe encourages physicalization and sonification designers to create and evaluate annotations early in the design process. We contend that this annotation-centric approach is necessary given the lack of formal conventions for sonifications and physicalization. Building upon Rahman et al.'s~\cite{rahman2023qualitative} taxonomy and design space of visual annotative techniques for visualizations, we explore how these visual techniques can be adapted to sonic and physical modalities. Through our design process, we found that annotations for these emergent representations require a broader interpretation than their traditional use in visualization. For instance, we identified modality-specific annotative techniques (e.g., timbre for sonification; moving the viewer's body to change frame of reference for physicalizations).
We evaluated \techname\ in an in-person workshop with designers ($N = 13$), which included reflective discussions on the purpose and design of annotations.

This work has two contributions. First, we explore how Rahman et al's~\cite{rahman2023qualitative} visual annotation design space can be extended for sonifications and physicalizations. Second, from analyzing the workshop's results, we found that annotations function beyond the conventional purpose of how to ``read'' a data representation---in these emergent data representations, they are inseparable from the data encodings and how people experience them. These insights suggest how sonification and physicalization designers might move from designing data representations to designing \textit{data experiences.}

\section{Related Work}

\subsection{Annotative Techniques in Visualization}
Data annotations are external visual marks intended to direct viewers' attention to important elements in a visualization ~\cite{rahman2023qualitative, matzen2017patterns,cedilnik2000procedural}. 
However, research on data annotations has focused on conventional visualizations, resulting in a research gap on how annotations should be integrated into other sensory data representations such as physicalizations and sonifications.

To address this gap, we focus on applying existing insights on visual annotations to sonifications and physicalizations. Given how physicalizations and sonifications lack formalized conventions, we contend that annotations are critical to communicating the data encoding process behind these emergent data representations.
Our work builds upon the taxonomy and design space proposed by Rahman et al.~\cite{rahman2023qualitative}, which describes seven annotation techniques: text, enclosure, connector, mark, color, trend, and geometric (\autoref{tab:annotation}).
We describe analogous annotation methods for sonifications and physicalizations by examining examples and reflecting on our past design practices. See \autoref{sec3:design-process} for more details on this process.

\subsection{Cards as Playful and Collaborative Design Tools}
Cards are popular analogue tools that serve as compact, tangible inspiration during the ``fuzzy front-end'' stages of design processes \cite{peters2021}. The use of cards in visualization research has mainly focused on pedagogy and teaching. For example, ViZItCards~\cite{he2016v} is intended to help students practice and reinforce visualization concepts. Cards have also been used in workshop settings to help generate ideas for physicalizations~\cite{huron2017let}. Worksheets are similar to cards as they generally focus on guiding novices or learners with best practices in visualization~\cite{mckenna2017worksheets-worksheets, wang2020cheat}. Cards are particularly well-suited for collaborative, playful ideation 
\cite{peters2021}, thus we aim to use them as a tool to enable exploratory ideation in a group setting. We developed cards specific to designing annotative techniques for physicalization and sonification design, which stand to benefit from the embodied, social nature of analogue design tools.

\section{\techname: An Annotation Game}
Using a RTD methodology~\cite{2007zimmerman, sedlmair2012design} to explore how data sonification and physicalization designers might more deliberately employ annotation techniques, we created \textit{\techname}, a design probe in the form of a collaborative, deck-based game. The probe's goal is for players to create and evaluate annotations for data sonifications or physicalizations that meet specific stakeholders' needs. 

\subsection{Game Design Process}
\label{sec3:design-process}
Our design process began with the authors performing close readings of sonification and physicalization examples hosted on archival websites~\cite{dataphys-wiki, sonification-archive} and survey papers~\cite{bae2022making, dragicevic2020data, hermann2011sonification} to identify their annotative techniques. Additional examples~\cite{perpetual-plastic, spiders-song} were added to capture relevant examples based on the authors' expertise. Using Rahman et al's~\cite{rahman2023qualitative} taxonomy as a starting point, we then focused on specific examples and discussed how annotative techniques are employed differently in sonic- and physical-tactile-media than in visual-only media (summarized in Table \ref{tab:annotation}).
We present two case studies to illustrate this process, highlighting how the \textit{geometric} visual annotation strategy is applied (see \autoref{fig:teaser}). See supplemental materials for more examples of different annotation techniques~\cite{baesupplmaterials}.

\textbf{``Spiders Song (Part 2)''}~\cite{spiders-song} is a sonification---conveying the evolution and phylogeny of spiders---embedded in an audio podcast. In the podcast, the hosts excerpt short audio clips from a longer sonification and discuss them in detail to emphasize specific information. Each short audio clip is a \textit{geometric} annotation because it presents a new perspective on a subset of the sonification. The podcast interweaves these short clips with a spoken narrative that \textit{connects} the clips with their descriptions (\autoref{fig:teaser}).

\textbf{``Perpetual Plastic''}
 \cite{perpetual-plastic} is an ephemeral art installation using plastic debris the artists collected on a beach to create a spiral 14-meters in diameter representing human plastic consumption.  
The physical installation contained no explicit annotative elements. However, in documenting the sculpture, the artists juxtaposed a zoomed-in view with a full-sculpture image, using a geometric annotation to recreate the experience of walking around the sculpture.
Photos documenting the installation have recreated this effect with aerial images and close-ups, augmented with \textit{text} labels.

\begin{figure}[t!]
  \centering
  \includegraphics[width=0.9\linewidth]{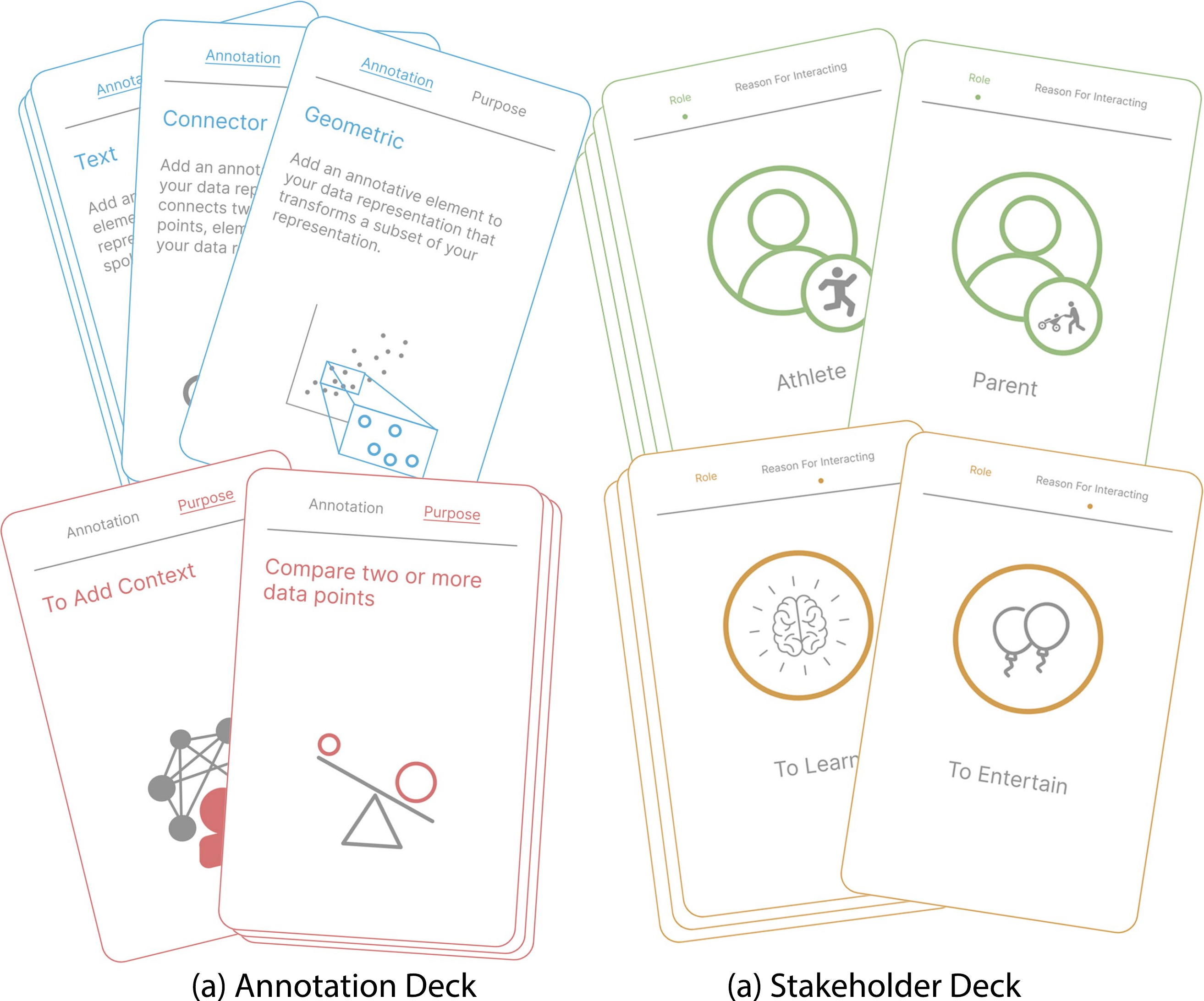}
  \caption{Two card sets in \techname: (a) The annotation cards highlight the different techniques (e.g., geometric, connector, text) as well as the purposes, such as adding context or comparing data points. (b) The stakeholder cards include different roles (e.g., parent, athlete) users might embody as well as reasons they might engage with a data representation (e.g., learn, entertain).}
  \label{fig:tk-name}
  \vspace{-1em}
\end{figure}

In both of these case studies, just as with visualizations, these annotative strategies occurred in \textit{ensembles} (i.e., combinations of techniques) \cite{rahman2023qualitative}. This observation inspired us to pursue cards, which lend themselves to brainstorming combinations of techniques. Further, our case studies encouraged us to focus on the purpose of annotations as well as their implementation. For example, while color does not have a direct sonic analog, its intended purpose is ``to highlight a specific data point or set of data points'' \cite{rahman2023qualitative}. Sonic-only features, like timbre (i.e., 
salient characteristic of a musical sound; a violin vs. a guitar) or volume can analogously represent this purpose; with physicalizations, this could be achieved with tactile-only features, like texture or temperature. This observation inspired us to create two different types of annotation cards: annotative techniques and the  explanatory purposes they might serve (\autoref{fig:tk-name}a). We also noted new techniques not included in the visual design space---like the use of repetition in sonifications---and incorporated these. 
We found that the cards supported our ideation processes, but required more scaffolding to generate actionable annotation designs. Given that annotation is ultimately about communication, we wanted to situate our designs within a playful, social context. Thus, we chose to create a game, meant for data designers to play together, to can support reflective dialog around annotation techniques.

\subsection{Game Components}
\label{sec3:game}
The game includes two card decks, a game board, worksheets, and instructions. All materials are included in the supplementary materials~\cite{baesupplmaterials}.

\textbf{Cards.} The game includes two card sets (\autoref{fig:tk-name}). The first set focuses on annotative \textit{techniques} (e.g., geometric, connector) and their \textit{purposes} (e.g., to add context; to confirm a user's understanding) (\autoref{fig:tk-name}a). The cards in this deck are drawn from the updated annotation taxonomy (\autoref{tab:annotation}) and are intended to scaffold designers on how to generate annotative elements for sonifications and physicalizations.
The second set focuses on stakeholders, and also has two types: the first type highlights different stakeholder \textit{roles} (e.g., teacher, medical professional) and their \textit{goals} (e.g., for personal growth, to learn) (\autoref{fig:tk-name}b). This cardset is intended to help designers tailor annotation ideas to specific stakeholders' needs.

\begin{figure}[tb]
  \centering
  \includegraphics[width=0.35\textwidth]{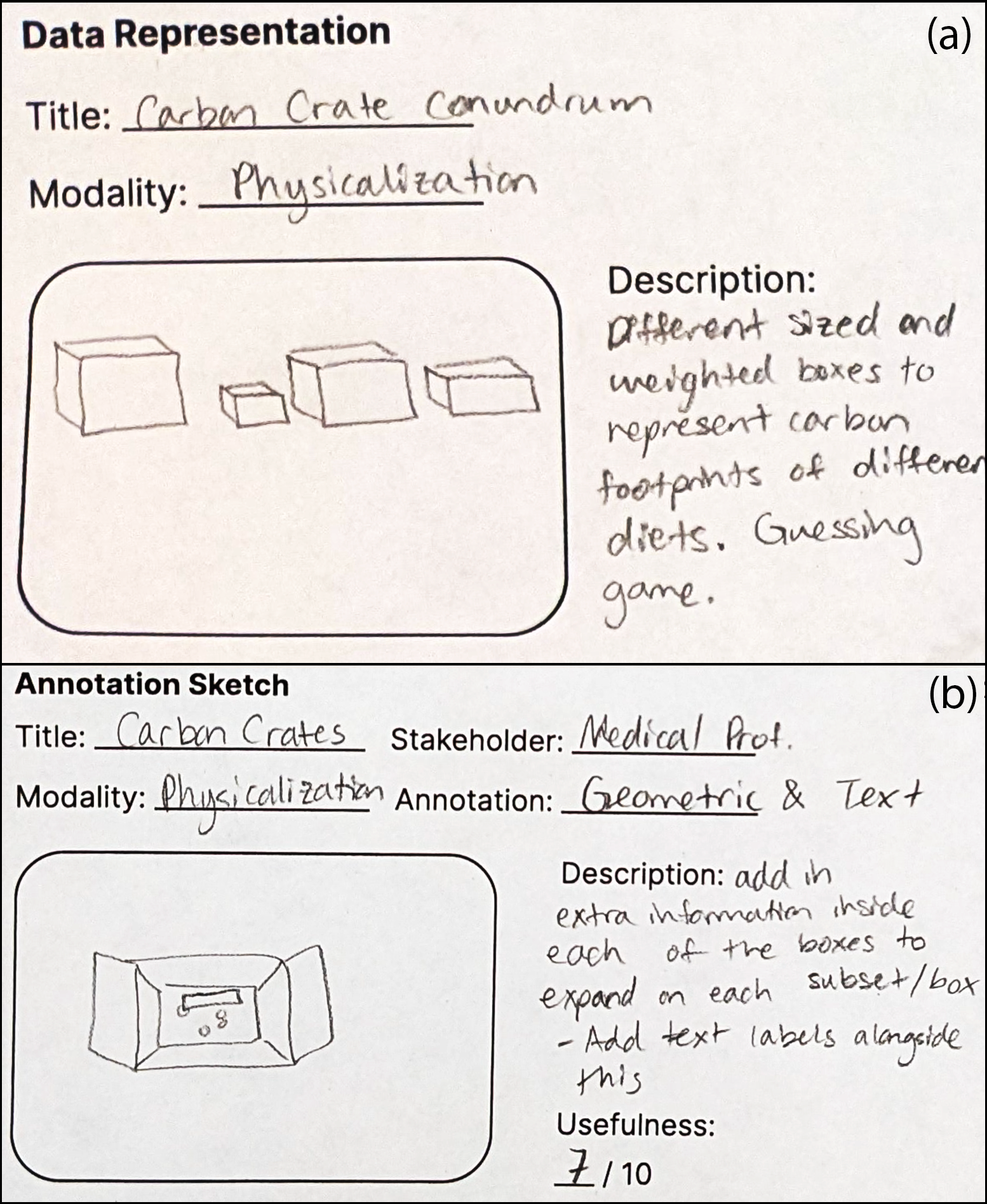}  \caption{Two worksheets filled out by participant (P1). (a) ``Data Representation'' describes a physicalization where boxes of different weights represent carbon footprints of various foods. (b) ``Annotation Sketch'' where P1 enhanced the physicalization with a \textit{geometric} annotation: Opening up a box reveals new information to add context.}
  \label{fig:worksheets}
  \vspace{-1em}
\end{figure}

\textbf{Game Board \& Worksheets.}
The game includes two different blank worksheet templates for players to document their design processes, drawing on the tradition of worksheets in visualization design \cite{fivedesignsheets}. The first worksheet is the data representation sheet, where players can write down the details of the sonifications or physicalizations they are designing. The second is an annotation sheet to display the annotative elements players generate as they play the game. See \autoref{fig:worksheets} for an example. Finally, the game includes a board for organizing the game elements and structuring the flow of gameplay, which can also serve as a standalone worksheet.

\textbf{Instructions.}
The goal of the game is to generate annotations for a data sonification or physicalization that meets a stakeholder's needs.
Here we provide a summary of the gameplay, and full instructions are detailed in the supplemental materials.

At the beginning of the game, two players are partnered, and each player engages in three rounds of activities.
In the first round (set up), each player fills out a blank data representation worksheet. They write a short description and sketch \cite{sensory-sketching} of the data physicalization or sonification they are working on (\autoref{fig:worksheets}a). Each player then draws a stakeholder card (\autoref{fig:tk-name}b), which is their identity for the rest of the game, and reveals it to the other player.
In the second round (design round), each player generates annotations for their data representation while addressing the stakeholder's needs that their partner is role-playing (\autoref{fig:worksheets}b).
Players can document their annotation ideas on the blank design sheets, putting one annotation idea on each sheet. Players are encouraged to generate as many annotations as they can within 10 minutes and to use the annotation deck (\autoref{fig:tk-name}a) as a creative brainstorming tool. 
n the third round (evaluation round), both players exchange their annotation ideas. They each evaluate the different annotation ideas from the perspective of the stakeholder that their partner is role-playing.

\section{Exploratory Study}
We conducted an exploratory study using \techname\ in an in-person workshop. Workshops can elicit rich qualitative insights for early stages of applied visualization research~\cite{kerzner2019workshops}. Using guidelines by Kerzner et al.~\cite{kerzner2019workshops}, we designed our workshop around the constraints of working with participants in a limited timeframe while still evaluating our research provocation. 
All workshop activities were approved by Whitman College's IRB. Study materials are available in supplemental materials~\cite{baesupplmaterials}. 

\subsection{Workshop}
\textbf{Participants.} We recruited undergraduate design students as our participants ($N= 13$) who were taking a ``Data Visceralization'' course, as they would be familiar with the practices of sonification and physicalization design. For the workshop, we asked participants to bring a data sonification or physicalization that they had either designed themselves or had previously analyzed in the course. Participants were compensated for their time with a \$10 gift card.

\textbf{Procedure.}
The 90-minute workshop was divided into three stages. The first stage consisted of an overview of the workshop and a reflective activity on the role of data annotations. The second stage focused on playing \techname\ for 60 minutes, following the gameplay outlined in \autoref{sec3:game}. The last stage was for reflecting on the workshop activities through group discussion and survey responses.

\subsection{Feedback from the Workshop}
We qualitatively analyzed materials from our exploratory study---including the artifacts participants generated while playing \techname\ (i.e., sketches), audio recordings of group discussions about participants' experiences of playing the game, and a written survey for participant feedback---using affinity diagramming as an iterative and inductive method.

A core theme we identified was how participants conceived of the purpose of annotations beyond traditional definitions that focus on explaining how audiences should interpret data representations. Participants found annotations for physicalizations and sonifications, \textit{``Bridge [the] gap between creator and audience'' (P2)} by conveying the story of its creation. Participants also noted that annotations \textit{``might serve as dialogue'' (P3)}: annotations can connect readers with each other, allowing them to contribute their own interpretations of a data representation.

Participants also noted that the game---and focusing on annotations and stakeholders---helped them `\textit{`think a bit more [creatively given the constraints], and help clarify my ideas better''} (P5). For example, P1 noted that the game caused them to realize that they had been integrating annotations into their design process without being conscious of it: \textit{``I had already made things I didn't realize were annotations.''} P6 noted that, \textit{``Annotations feel like a part of a physicalization/sonification, rather than just a helping guide.''} 

Participants also experienced challenges associated with coming up with annotations for sonifications and physicalizations. This challenge was especially prevalent for sonifications: \textit{``[it's] difficult to translate some annotative techniques across modalities...what is gestalt in sound?'' (P4).} In fact, upon reflection, participants noted that the process of coming up with new annotation strategies might mean completely re-designing a sonification or physicalization.
For example, P1 considered re-designing a physicalization by incorporating the gesture of opening a box that contains the physicalization to provide contextual information about the data encoding (\autoref{fig:worksheets}b). And
P7 considered re-designing a sonification into a pop-song where the lyrics can be an example of text annotations. This suggests that participants found annotations and data encodings to be tightly coupled. To address these challenges and facilitate the process of generating annotations, participants suggested providing a gallery of non-visual annotation examples.

\section{Future Work and Discussion}
In this exploratory design study, we considered how focusing on annotations might help us create more understandable data representations. We advocate treating annotation as a holistic, integrated element when designing physicalizations and sonifications, rather than an addendum at the end of the design process. By doing so, we can identify new and compelling ways to encode data and communicate those encodings to audiences. This perspective shift opens future research directions to (i) develop further annotation-centric tools to support sonification and physicalization designers, (ii) consider how the field of visualization can draw from annotative practices in sonification and physicalization, and (iii) explore annotation as a practice that can facilitate two-way communication between designers and end-users.

\textbf{Developing annotation-centric tools.} In the future, we hope to refine \techname\ and to develop additional features to support designers in thinking about annotations. For example, some participants found having an assigned stakeholder helped them generate ideas, whereas others found it too constraining. We hope to extend the game to offer different levels of creative constraints, such as offering different ``modes'' of play, like collaborative mode or competitive mode. Drawing inspiration from Bach et al's work on data comics~\cite{bach2018}, we also hope to provide examples of the different annotative techniques used in sonifications and physicalizations in the form of a gallery or library of design patterns \cite{data-comics-website}. We hope such tools can support data designers in creating their own, personal, annotation first-design practice.

\textbf{Exploring design spaces across sensory modalities.} Enge et al's~\cite{enge_towards_2023} recent theoretical work presents a unified design space between visualizations and sonifications, which demonstrates that a shared, multimodal design language can lead to novel ideas for representing data. We extend this line of work by considering the role of cross-sensory annotations in helping users learn how to interpret data representations. Our exploratory design study highlights how sonification and physicalization designers can leverage existing visualization insights. We found that focusing on annotative techniques drawn from visualization generated new ideas for \textit{encoding} data in sound and physical objects, as well as for annotating them. This suggests that visualization designers might, likewise, be able to learn something new from considering the unique, modality-specific annotative strategies that sonification and physicalization designers use. For example, the use of repetition in sonification as an annotative technique might offer new ways to design time-series visualizations~\cite{fang2020survey}. Gesture to make sense of physicalizations could suggest embodied techniques for visual data storytelling~\cite{tong2018storytelling, gershon2001storytelling}. 

This paper is non-exhaustive. In this paper, we used a case-study approach (\autoref{sec3:design-process}), but in the future, we hope to do a more comprehensive analysis of the examples that are available in public repositories such as the Sonification Archive \cite{sonification-archive} and the Physicalization Wiki \cite{dataphys-wiki}. We hope to identify additional annotative techniques that are unique to sonification and physicalizations, with the ultimate goal of translating design knowledge across sensory modalities.

\textbf{Exploring annotation as a social practice.} Annotations allow designers to communicate information---including data encodings, context, and intent---to end users. However, our participants had a broader view of annotations. They viewed annotations as a tool to provide stakeholders to share their interpretations back to the designer as well as with other stakeholders. This insight mirrors the Friske et al.'s~\cite{friske2020} findings, which highlight how people \textit{experiencing} a data representation might result in re-creating their own interpretations. Future work can explore how we might capture and share annotations as a social practice contributed by end-users. This also encourages us to pursue new ways to evaluate the effectiveness of annotations, given the multiple purposes they serve.

Taken together, we hope these future directions might encourage data designers to more deliberately consider end users and move beyond designing data artifacts into designing \textit{data experiences}.

\section{Conclusion}
We explored the role of annotations in the sonification and physicalization design process through a multi-faceted, research-through-design process. We created a design probe, in the form of a card-based game, to support sonification and physicalization designers in foregrounding annotations and play-tested it in a participatory workshop. Through this process, we demonstrated how annotations are an integral, holistic element of sonifications and physicalizations and how focusing on annotation design can surface new ideas for data encodings.

\acknowledgments{
We would like to thank our colleagues and anonymous reviewers who helped \textit{shape} this work and contributed to its \textit{soundness}.
}

\bibliographystyle{abbrv-doi-hyperref}
\bibliography{bib}
\end{document}